\documentclass[aps,showpacs,onecolumn,
superscriptaddress]{revtex4}
\usepackage{graphicx}
\usepackage{dcolumn}
\usepackage{bm}
\usepackage{color}
\usepackage[normalem]{ulem} 
\usepackage[dvipsnames]{xcolor} 
\usepackage{hyperref}
\hypersetup{
  colorlinks=true,        
  linkcolor=blue,         
  citecolor=cyan,         
}
\usepackage{mathrsfs}
\usepackage[utf8]{inputenc}
\usepackage{mathtools}
\usepackage{doi}
\usepackage{amsmath}
\usepackage{amssymb}

\begin{document}

\title{Dynamics of particles with electric charge and magnetic dipole moment near Schwarzschild-MOG black hole}

\author{Sardor Murodov} \email{mursardor@gmail.com}
\affiliation{Institute of Fundamental and Applied Research, National Research University TIIAME, Kori Niyoziy 39, Tashkent 100000, Uzbekistan;}
\affiliation{Uzbekistan-Finland Pedagogical Institute, Spitamen shokh str. 166, Samarkand 140100, Uzbekistan;} \affiliation{Department of Theoretical Physics, Samarkand State University, Samarkand 140104, Uzbekistan;}

\author{Javlon Rayimbaev}
\email{javlon@astrin.uz}
\affiliation{School of Mathematics and Natural Sciences, New Uzbekistan University, Mustaqillik Ave. 54, Tashkent 100007, Uzbekistan;}
\affiliation{School of Engineering, Central Asian University, Tashkent 111221, Uzbekistan;}
\affiliation{Institute of Fundamental and Applied Research, National Research University TIIAME, Kori Niyoziy 39, Tashkent 100000, Uzbekistan;}

\author{Bobomurat Ahmedov}
\email{ahmedov@astrin.uz}
\affiliation{Institute of Fundamental and Applied Research, National Research University TIIAME, Kori Niyoziy 39, Tashkent 100000, Uzbekistan;}
\affiliation{Ulugh Beg Astronomical Institute, Astronomy St 33, Tashkent 100052, Uzbekistan} 
\affiliation{Institute of Theoretical Physics, National University of Uzbekistan, Tashkent 100174, Uzbekistan;}

\author{Abdullo Hakimov}
\affiliation{Samarkand International University of Technology, 270 Spitamen Avenue, Samarkand, Uzbekistan}

\date{\today}

\begin{abstract}
Investigations of electromagnetic interactions between test-charged and magnetized particles are important in the dynamics of the particles in strong gravitational fields around black holes. Here, we study the dynamics of a particle having an electric charge and a magnetic dipole moment in the spacetime of a Schwarzschild black hole in modified gravity (MOG), called Schwarzschild-MOG black hole. First, we provide a solution of Maxwell equations for the angular component of electromagnetic four potentials in the Schwarzschild-MOG spacetime. Then, we derive equations of motion and effective potential for circular motion of such particles using a hybrid form of the Hamilton-Jacobi equation which includes both interactions of electric charge and magnetic dipole moment with the external magnetic field assumed as asymptotically uniform, and interaction between the particles and the MOG field. Also, we investigate the effects of the three types of interactions on the radius of innermost stable circular orbits (ISCOs) and the energy \& angular momentum of the particles at their corresponding ISCOs. Finally, we provide detailed analyses of the effects of the three interactions mentioned above on the center of mass energy in the collisions between neutral, electrically charged, and magnetized particles.

\end{abstract}

\maketitle

\section{Introduction}

The present theory and observations confirm that our Universe consists of 73 \% dark energy, 23 \% dark matter, and the rest 4 \% ordinary matter ~\cite{2009Natur.458..587C,CoGeNT:2010ols}. However, there is no unique and complete theory that describes dark matter and dark energy (gravitational) physics. In fact, general relativity (GR) is a classical gravitational theory that is well-tested in weak gravitational field regimes and at the moment started to be experimentally and observationally verified in the strong field regime. Unfortunately, the GR as classical theory has a singularity problem that the theory itself can not explain or avoid. However, mathematically, the problem can be avoided by modifying the GR with non-linear electrodynamic, scalar, and quantum fields. 

In particular, J. Moffat ~\cite{Moffat06} proposed a new approach to modify the GR that promises models of gravity that may build the unified theory of gravity. Actually, it contains a massive scalar field, which is referred to as scalar-tensor-vector gravity (STVG).

In order to avoid the break-down of the spacetime around black holes at short distances in the frame of modified gravity (MOG) theory in terms of the massive vector field which has a source scalar charge $Q=\sqrt{\alpha G}M$, where $G$ is the Newtonian constant, $M$ is the total mass of the black hole and $\alpha$ is coupling parameter so-called MOG parameter. This term has an additional force with a repulsive nature and is significant at the quantum level. There are several black hole solutions that have been obtained in MOG gravity.

For instance, non-rotating and rotating black hole solutions have been obtained in Ref.~\cite{Moffat15}. Effects of the STVG field on the dynamics of test particles and stability of their circular orbits around Schwarzschild-MOG black holes have been explored in the presence and absence of external magnetic fields~\cite{Hussain15,Boboqambarova2023MPLA...3850071B}. Solar system tests have also been successfully analyzed in Ref~\cite{Moffat06}, galaxy rotation curve~\cite{Moffat13,Moffat15a}, testing MOG using data from X-ray observations of galaxies~\cite{Moffat14} and S2 star motion \cite{Turimov2022MNRAS.516..434T,Della2022MNRAS.510.4757D}, shadow of black holes ~\cite{Moffat15,Moffat15b}, thermodynamic properties ~\cite{Mureika16}, supernovae ~\cite{Wondrak18}, gravitational lensing ~\cite{Moffat09}, quasinormal modes ~\cite{Manfredi18}, gravitational waves in MOG \cite{2016PhRvD..94l4038D,2021arXiv211107704C} and electromagnetic fields around neutron stars \cite{Rayimbaev2020PhRvD.102b} have been extensively studied. The different properties of the spacetime around black holes in MOG have been explored in Refs.~\cite{Pradhan18,Pradhan19,Kolos20,Sharif17,Shojai17,Haydarov2020EPJC,Rayimbaev2023EPJP..138..358R,Rayimbaev2021Galax...9...75R,Turimov2023PhLB..84338040T}.

The electrically charged and magnetized particle dynamics have been well studied in the spacetime of different black holes immersed in external magnetic fields.

In our recent work \cite{Rayimbaev2023Univ....9..135R}, we have investigated the dynamics of a test particle with the electric charge and magnetic dipole moment around a magnetized Schwrazschild black hole. It is shown that slowly rotating magnetized 
 neutron stars and white dwarfs can be interpreted as such particles (the interaction between spin of the such astrophysical objects and the curved spacetime is small enough to neglect it).

In the present work, we aimed to study the motion of charged particles with dipole moment around Schwraschild black holes in MOG taking into account the interaction between particles and the STVG field. The paper is organized as follows. In section \ref{Sec:PM}, we briefly introduce spherically symmetric (static) black hole solution in modified gravity and describe external magnetic fields around the black holes. Section \ref{partmotion} is devoted to deriving effective potential for the circular motion of the particles and investigating their ISCOs. In Section \ref{Sec:Collision} we study critical angular momentum and center of mass energy of charged, neutral, and magnetized particles. We summarize the obtained results in Section \ref{sec:6}. 

Throughout the paper, we use geometrized units $c=G=1$ and run the Latin indexes from 0 to 3 and Greek ones from 1 to 3.

\section{Schwarzschild black holes in MOG}\label{Sec:PM}
{The gravitational field action in the STVG theory includes GR $S_{G}$, matter (pressure-less) $S_{M}$, vector field $S_{\phi}$ and scalar field $S_{S}$ terms \cite{Mureika16}:
\begin{equation}
    \label{action}
 S = S_{G}+S_{\phi}+S_{S}+S_{M}\ ,
\end{equation}
where 
\begin{eqnarray}
&& S_{G} = \frac{1}{16\pi}\int \frac{1}{G} (R+2 \Lambda) \sqrt{-g}d^4x,
\\ && S_{\phi} = -\frac{1}{4\pi} \int \left[{\cal K} + {\rm V}(\phi)\right] \sqrt{-g}d^4x\ , 
\\
&& S_{S} = \int \frac{1}{G} \Bigg[\frac{1}{2} g^{\alpha \beta} \left( \frac{\nabla_{\alpha}G \nabla_{\beta}G}{G^2} + \frac{\nabla_{\alpha}\mu \nabla_{\beta}\mu}{\mu2} \right)- \frac{V_{G}(G)}{G^2} -\frac{V_{\mu}(\mu)}{\mu^2} \Bigg] \sqrt{-g}d^4 x\ ,
\\
&&  S_{M} = -\int (\rho \sqrt{u^{\mu}u_{\mu}}+{\cal Q}u^{\mu}\phi_{\mu})\sqrt{-g}d^4 x + J^{\mu} \phi_{\mu}\ ,
\end{eqnarray}
with $R = g^{\mu\nu} R_{\mu\nu}$ is the Ricci scalar, $\Lambda$ is the cosmological constant, $g\equiv$det$(g_{\mu\nu})$ is the determinant of the metric tensor, $\nabla_{\alpha}$ is the covariant derivation, ${\cal K}$ is the kinetic term for the scalar field $\phi_{\mu}$ which reads as 
\begin{equation}
    4 {\cal K} = B^{\mu\nu} B_{\mu\nu}\ ,
\end{equation}
where $B^{\mu\nu} = \partial _{\mu} \phi_{\nu} -\partial _{\nu} \phi_{\mu}.$ The covariant current density is defined to be 
\begin{equation}
 J^{\mu} =\kappa T_M^{\mu\nu} u_{\nu}\ ,
 \label{current-dens}
\end{equation}
where $T_M^{\mu\nu}$ is the energy-momentum tensor for matter with $\kappa = \sqrt{\alpha G_N},$ $\alpha = (G-G_N) /G_N$ is a parameter defining the scalar field, $G_N$ is Newtonian gravitational constant, $u^{\mu} = dx^{\mu} /d\tau$ is a timelike velocity and $\tau$ is the proper time a long time like geodesic. The perfect fluid energy-momentum tensor for matter is given by 
\begin{equation}
 T^{M\mu\nu} = (\rho_M + p_M) u^{\mu}u^{\nu} -p_Mg^{\mu\nu}\ ,
 \label{tensor-energy}
\end{equation}
where $\rho_M$ and $p_M$ are the density and pressure of matter, respectively. 
From Eqs. (\ref{current-dens}) and (\ref{tensor-energy}) using $u_\mu u^\mu=1$, we get 
\begin{equation}
 J^{\mu}={k} \rho_M u^{\mu}\ . 
\end{equation}
For the matter-free and pressureless MOG field ($T_M^{\mu\nu}=0$) in the asymptotically flat (zero-cosmological constant) spacetime, the field equation takes the form 
\begin{equation}
    G_{\mu\nu}=-\frac{8\pi G}{c^4}T^\phi_{\mu\nu}\ ,
\end{equation}
where $T^\phi_{\mu\nu}$ is the tensor of massive-vector field. The observational data from galaxy and cluster dynamics show that the mass of the particles of the field $\phi$ is about $m_\phi=2.6\times10^{-28}$ eV, and it is almost zero \cite{Moffat13}. One may assume that the vector field is an analogue of the electromagnetic field, and its field tensor is defined as
\begin{equation}
    T^\phi_{\mu\nu}=-\frac{1}{4\pi}(B_\mu^\alpha B_{\nu \alpha} - \frac{1}{4} g_{\mu \nu} B^{\alpha\beta}B_{\alpha\beta})
\end{equation}
with 
\begin{eqnarray}
 && \Delta_\mu B^{\mu\nu}=0\ , \\ && \Delta_\alpha B^{\mu\nu}+\Delta_\nu  B^{\mu\alpha}+\Delta_\mu B^{\alpha\nu}=0\ .
\end{eqnarray}
The above assumptions imply that the potential term of the action $S_\phi$ is zero (${\rm V}(\phi)=(1/2)\mu \phi_\mu\phi^\mu=0$), so it has only kinetic term, and one may consider the kinetic term is a function of the massive-vector field invariant ${\cal B}=B_{\mu\nu}B^{\mu\nu}$ as ${\cal K}=f({\cal B})$. 

The spacetime of Schwarzschild black hole in MOG is described by the line element 
\begin{equation}\label{metric}
ds^2=-\frac{\Delta(r)}{r^2}dt^2+\frac{r^2}{\Delta(r)}dr^2+r^2(d\theta^2+\sin^2\theta d\phi^2)\ ,
\end{equation}
where 
\begin{equation}
\Delta(r)=r^2-2 (1+\alpha) M r+\alpha (1+\alpha)M^2 \ .
\end{equation}

\subsection{Magnetization of Schwarzschild-MOG black holes}

In realistic astrophysical cases, magnetic field configurations near black holes are very complex due to ionized accretion disc dynamics.  

The simple approach for getting analytical expression for the magnetic field around the Schwarzschild black hole is Wald's one \cite{Wald74} and the following exact analytical expression for the electromagnetic vector potential 
\begin{equation}
\label{mag21}
A_{\phi} =\frac{1}{2}B_0r^2\sin^2\theta \ 
\end{equation}
is derived in the spacetime of the Schwarzschild black hole assuming that the black hole is immersed in an external asymptotically uniform magnetic field 
where $B_0$ is the asymptotic value of the magnetic field. It is worse to note that Eq. (\ref{mag21}) satisfies the Maxwell equations 
\begin{equation}
\label{mag22}
    \frac{1}{\sqrt{-g}}\partial_\mu (\sqrt{-g}F^{\mu\nu})=0 , \quad F_{\mu\nu}=A_{\nu,\mu}-A_{\mu,\nu} \ . 
\end{equation}

For example in Ref. \cite{Boboqambarova2023MPLA...3850071B}, the expression for the vector potential has been taken  in the standard form 
\begin{equation}
\label{mag23}
A_{\phi} = \frac{1}{2}B_0 \psi(r)\sin^2\theta \ ,
\end{equation}
where $\psi(r) $ is a radial function found as a solution of Maxwell equation (\ref{mag22}). In the GR limit at $\alpha=0$, the function $\psi(r)=r^2$.

The exact analytical solution of the Maxwell equation for $A_\phi$ in the Schwarzschild-MOG black hole spacetime is found by inserting Eq. (\ref{mag23}) into Eq. (\ref{mag22}), in the following form:

\begin{equation}
\label{mag25}
A_{\phi}=\frac{1}{2}B_0\left[r^2-\alpha(\alpha+1)M^2\right]\sin^2\theta \ .
\end{equation}

The magnetic field around the Schwarzschild BH, measured in the proper observer frame of reference, is 
\begin{equation}\label{fields}
B^{\alpha} = \frac{1}{2} \eta^{\alpha \beta \sigma \mu} F_{\beta \sigma} w_{\mu}\ ,
\end{equation}
where, $w_{\mu}$ is four-velocities of an observer, $\eta_{\alpha \beta \sigma \gamma}$  is the pseudo-tensorial form of the Levi-Civita symbol $\epsilon_{\alpha \beta \sigma \gamma}$ which has the form 
\begin{equation}
\eta_{\alpha \beta \sigma \gamma}=\sqrt{-g}\epsilon_{\alpha \beta \sigma \gamma}\ , \qquad \eta^{\alpha \beta \sigma \gamma}=-\frac{1}{\sqrt{-g}}\epsilon^{\alpha \beta \sigma \gamma}\ .
\end{equation}
 The Levi-Civita symbol is  $\epsilon_{0123}=1$ for even permutations, and it is -1 for odd permutations. Consequently, one can obtain the non-zero components of the external magnetic field measured by a proper observer with the four-velocities $w^{\mu}_{proper}=(1/\sqrt{f(r)},0,0,0)$  read
\begin{equation}
    B^{\hat{r}}=B_0 \left(1+\frac{\alpha(\alpha+1)M^2}{r^2}\right) \cos\theta, \ B^{\hat{\theta}}=\sqrt{f(r)}B_0\sin \theta\ .
\end{equation}

\section{Equations of motion for charged magnetized particles around Schwarzschild MOG black holes  \label{partmotion}}

The equations of motion of charged particles with magnetic dipole around magnetized Schwarzschild black holes in STVG using the hybrid form of the Hamilton-Jacobi equation taking into account interaction between the particle and STV field,
\begin{equation}\label{HJ}
g^{\mu \nu}\left(\frac{\partial {\cal S}}{\partial x^{\mu}}+ eA_\mu+\hat{q}{\Phi}_\mu\right) \left(\frac{\partial {\cal S}}{\partial x^{\nu}}+eA_{\nu}+\hat{q}{\Phi}_\nu\right)= -\Bigg(m-\frac{1}{2}U \Bigg)^2,
\end{equation}

 where $\hat{q}=\sqrt{\alpha}m$, $\Phi_\mu$ is the scalar potential of the massive scalar field, the term $U=D^{\mu \nu}F_{\mu \nu}$ is responsible for the interaction between the magnetic dipole and the external magnetic field.  $D^{\mu \nu}$ and $F_{\mu \nu}$ are polarization and electromagnetic field tensors, respectively. The expression for the tensor $D^{\mu \nu}$ has the form ~\cite{deFelice03}:
\begin{equation} \label{Dab}
D^{\alpha \beta}=\eta^{\alpha \beta \sigma \nu}u_{\sigma}\mu_{\nu} , \qquad D^{\alpha \beta }u_{\beta}=0\ ,
\end{equation}

where $\mu^{\nu}$ and $u^{\nu}$ respectively refer to the four-dipole moment vector and the four-velocity of the particle measured by a proper observer. 

The four-potential of the scalar field \cite{Turimov2022MNRAS.516..434T}
\begin{equation}
    \Phi_\mu=-\frac{\sqrt{\alpha}M}{r}\left(1, 0, 0, 0\right)\ . 
\end{equation}
The electromagnetic field tensor has the following expression in terms of electric $E_{\alpha}$ and magnetic $B^{\alpha}$ fields :
\begin{eqnarray}  F_{\alpha \beta}=u_{[\alpha}E_{\beta]}-\eta_{\alpha \beta \sigma \gamma}u^{\sigma}B^{\gamma}\, . \end{eqnarray}
%
%
Taking into account the condition given in Eq.~(\ref{Dab}) the product of polarization and electromagnetic tensors for the proper observer is defined by  
\begin{equation}\label{DF1}
U=2\mu^{\hat{\alpha}}B_{\hat{\alpha}}
 =2\mu B_0 \sqrt{f(r)}\, .
 \end{equation}

For further analysis, we assume the direction of the dipole moment has the components $\mu^{\alpha}=(0,\mu^{\theta},0)$ which is always parallel to the magnetic field lines and perpendicular to the equatorial plane. 

One may derive equations of motion using the Lagrangian taking into account MOG interaction \cite{Turimov2022MNRAS.516..434T}
\begin{equation}\label{v2}
    \mathscr{L}=\frac{1}{2}g_{\mu\nu}u^{\mu}u^{\nu}+qA_\mu u^\mu+\frac{\Tilde{q}}{m}\Phi_\mu u^\mu\ .
\end{equation}

The integrals of motion for electrically charged particles with magnetic dipole moment can be found using the time translation and the rotational symmetry of the geometry which corresponds to the conserved quantities that can be calculated using the Killing vectors
\begin{equation}
\xi_{(t)}^{\mu}\partial_{\mu}=\partial_{t} , \qquad
\xi_{(\phi)}^{\mu}\partial_{\mu}=\partial_{\phi},
\end{equation}
here
$\xi_{(t)}^{\mu}=(1,~0,~0,~0)$ and $\xi_{(\phi)}^{\mu}=(0,~0,~0,~1)
$, and the corresponding conserved quantities are the specific energy
${\cal E}=E/m$ of the moving
particle and its angular momentum ${\cal L}=L/m$. Consequently, $\dot{t}$ and $\dot{\phi}$ take the form

\begin{eqnarray}
\label{tdot}
\dot{t}&=&\frac{1}{f(r)} \left({\cal E}-\frac{\alpha M}{r} \right)\ ,
\\ \label{fdot}
\dot{\phi}&=&\frac{l}{r^2}-\omega\left[1-\alpha(1+\alpha)\frac{M^2}{r^2} \right]\ .
\end{eqnarray}

Accordingly, the dynamics of electrically charged particles with magnetic dipole moment orbiting the magnetized Scharzschild-MOG black hole at the equatorial plane (i.e., $\theta=\pi/2$ and $\dot{\theta}=0$) will be described by the following action:
\begin{equation}\label{action}
{\cal S}=-Et+L\phi +{\cal S}_r 
\ .
\end{equation}

Then we can derive the equation for the radial coordinate using the Hamilton-Jacobi equation (\ref{HJ}) and the obtained integrals of motions along $t$ and $\phi$ coordinates (\ref{tdot}) and (\ref{fdot}) as 
\begin{eqnarray}
    \label{rdot}\nonumber
\dot{r}^2=\left({\cal E}-\frac{\alpha M}{r} \right)^2-f(r)\Big[\left(1-\beta \sqrt{f(r)}\right)^2+\left(\frac{\mathcal{L}}{r}+\omega r\left(1-\alpha(1+\alpha) \frac{M^2}{r^2}\right)\right)^2\Big]\ .
\end{eqnarray}
 
With this in mind, one can write the variables in a separate form in the Hamilton-Jacobi equation. The radial motion of the particle can be then defined by  
\begin{equation}
\dot{r}^2={\cal{E}}^2-V_{\rm eff}(r)\, ,
\end{equation}
%
where $V_{\rm eff}(r)$ for the circular motion of charged magnetized particles has the form 
\begin{equation}\label{effpoeq}
V_{\rm eff}(r)=-\frac{\alpha M}{r}+\sqrt{f(r)\left[\left(1-\beta \sqrt{f(r)}\right)^2+\left(\frac{\mathcal{L}}{r}+\omega r \left(1-\alpha(1+\alpha) \frac{M^2}{r^2}\right)\right)^2\right] }\, ,
\end{equation}

where ${\cal L}=L/m$ is the specific angular momentum of the particles and $\beta = \mu B_0/(2m)$ is the magnetic coupling parameter corresponding to the magnetic interaction between the magnetic dipole moment of the particles and external magnetic fields, $\omega=eB_0/(2m)$ is the Larmor frequency parameter.

\begin{figure}[ht!]
\centering
\includegraphics[width=0.6\textwidth]{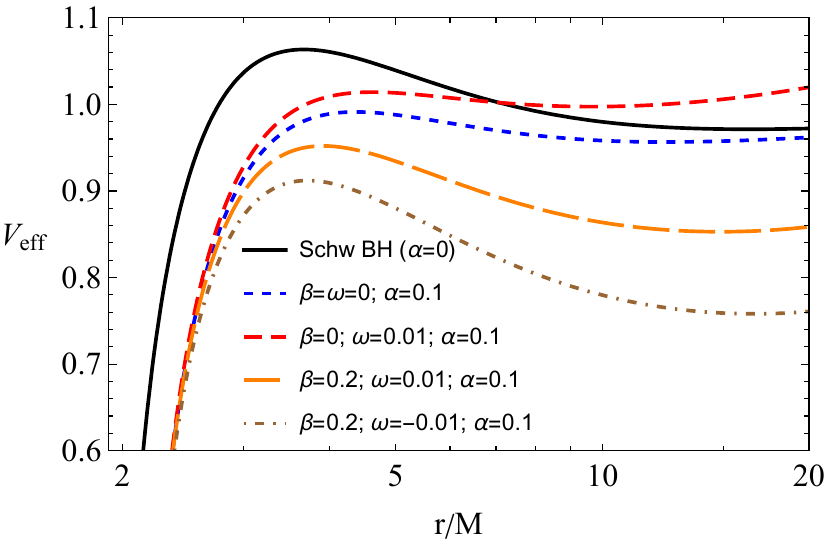}
\caption{Radial profiles of the effective potential for the circular motion of charged magnetized particles around Schwarzschild-MOG black holes for different values of $\alpha, \beta$, and $\omega$ parameters.\label{fig: Veff}}
\end{figure}

The radial dependence of the effective potential for the circular motion of charged particles with magnetic dipole moment around Schwarzchild black holes in modified gravity for different values of $\alpha, \beta$, and $\omega$ parameters with the comparisons of the Schwarzschild black hole case. It is observed that in the presence of the MOG field, the maximum of the effective potential decreases due to the negative interaction energy between test particles and the STV field. Also, the maximum increase (decrease) is due to the presence of an electromagnetic interaction between the electric charge and the magnetic field when $\omega>0$ ($\omega<0$). However, in the presence of magnetic interaction between the magnetic field and the magnetic dipole of the particles, the effective potential decreases sufficiently.

\subsection{Circular orbits}
Along circular orbits there is (no radial motion) no radial forces or the existing forces compensate each other at the corresponding values of angular momentum of the particles.

One can study the circularity of the orbits of test-charged magnetized particles orbiting the magnetized black hole using conditions $V_{\rm eff}={\cal E}$ and $ V_{\rm eff}'=0$, where the prime stands for the partial derivative with respect to the radial coordinate. Solving this condition, one can find the angular momentum of particles corresponding to circular orbits.

\begin{figure*}[ht!]
\centering
\includegraphics[width=0.6\textwidth]{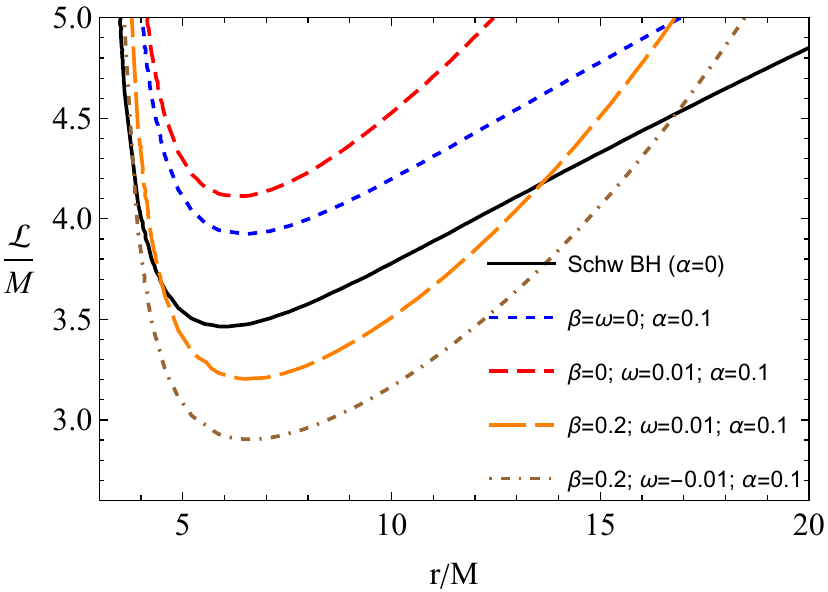}
\caption{Angular momentum of test-charged-magnetized particles orbiting Schwarzschild-MOG black holes for various values of $\alpha, \beta$, and $\omega$ parameters. \label{fig:AMom}}
\end{figure*}

Figure \ref{fig:AMom} shows the specific angular momentum of charged magnetized particles moving around the Schwarzschild black hole in modified gravity for different values of the parameters $\alpha, \beta$, and $\omega$. It is seen from the figure that the minimum in the angular momentum increases sufficiently in the presence of the MOG field parameter $\alpha$, and the distance where it is minimum goes far from the central object (see black solid and blue-dashed lines). Similarly, the angular momentum also increases due to the presence of positive values of the magnetic coupling parameter $\omega$. Meanwhile, when $\omega<0$ and the presence of magnetic interaction parameter $\beta$ causes a decrease in the minimum of the angular momentum.  

\subsection{Innermost stable circular orbits}

Solving condition $V_{\rm eff}'=0$ with respect to $r$ helps to find the orbits where the effective potential has extreme values. The circular orbits become stable where the effective potential is minimal. Thus, $V_{\rm eff}''(r)<0$ the orbits are unstable and all stable circular orbits satisfy condition $\partial_{rr}V_{\rm eff}(r_{\rm ISCO})>0$, while ISCO satisfies $\partial_{rr}V_{\rm eff}(r_{\rm ISCO})=0$. The importance of the ISCO around black holes is connected with the inner edge of the accretion disc.
Interestingly, when test particles in their Keplerian accretion disk, they fall down into the central black hole and extract some amount of energy which may convert to both electromagnetic and gravitational radiation under certain conditions. The energy released through the radiations can be determined by the difference between the rest of the energy of the particle (measured by a suitable observer) and the ISCO energy of the particles (${\cal E}_{\rm ISCO}$). Consequently, the efficiency of the energy release from the accretion disk has the following form \cite{Novikov73}
\begin{equation}
\eta=1-{\cal E}\vline_{\,r=r_{\rm ISCO}}\ . 
\end{equation}

Below, we analyze the effects of the STV field on the radius of ISCO of test particles, their energy and angular momentum at the orbits, and the energy efficiency for different values of magnetic coupling and magnetic interaction parameters between magnetic field and electric charge \& magnetic dipole moment of the particles, respectively.

\begin{figure*}[ht!]
\centering
\includegraphics[width=0.42\textwidth]{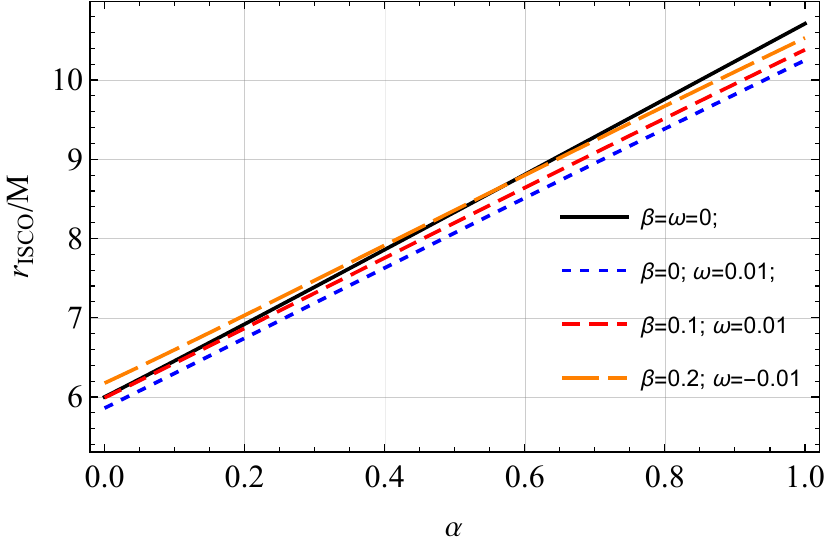}
\includegraphics[width=0.43\textwidth]{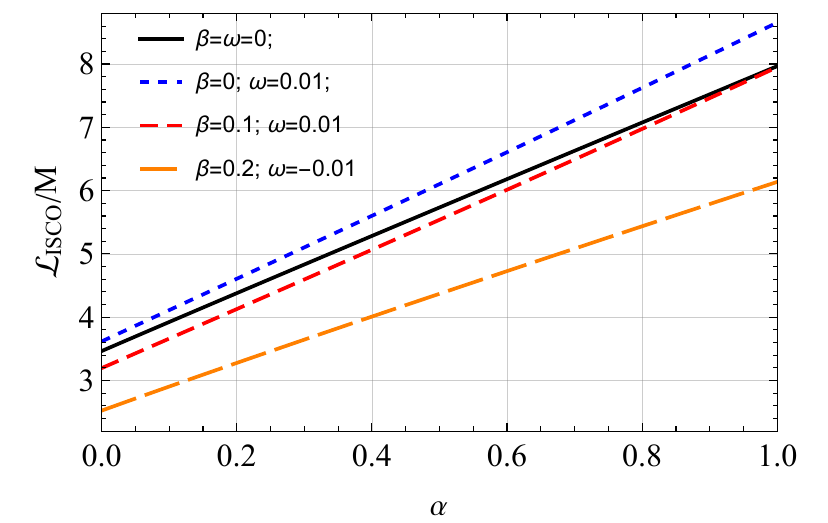}
\includegraphics[width=0.43\textwidth]{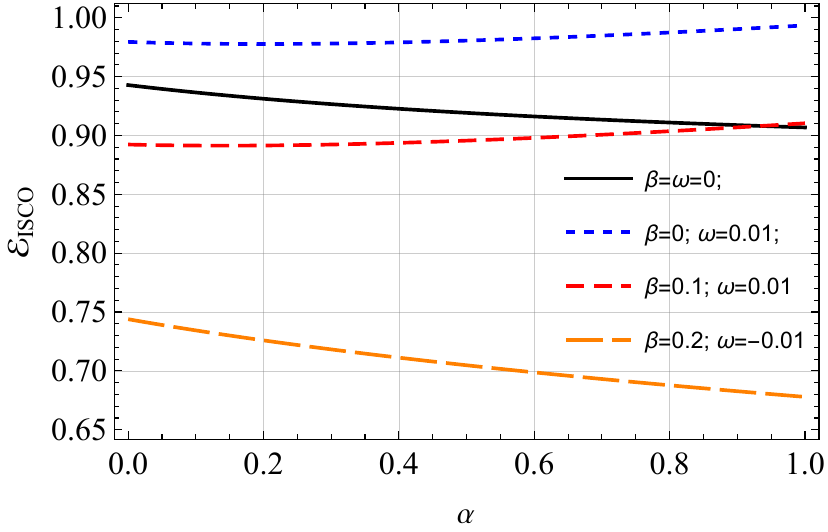}
\includegraphics[width=0.43\textwidth]{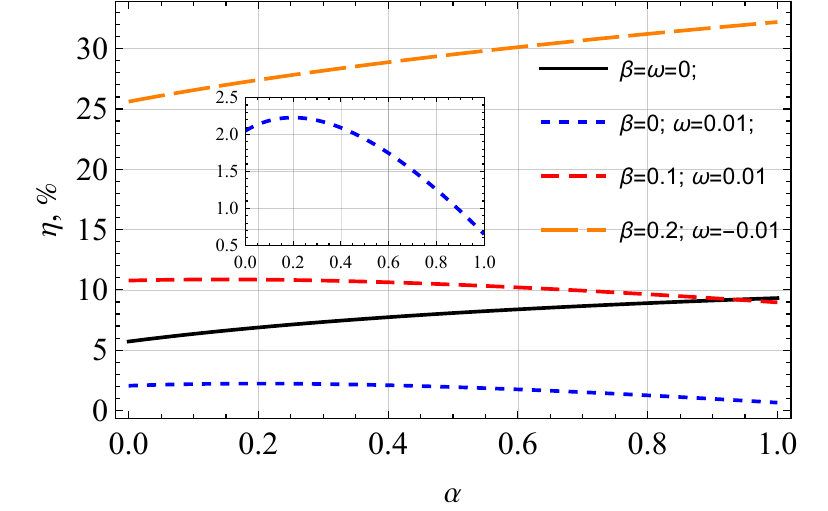}
\caption{The dependence of ISCO radius (top left), angular momentum (top right), and energy (bottom left) of the particles at ISCO and the energy efficiency,$\eta$ (bottom right) from the MOG field parameter $\alpha$.\label{fig:withalpha}}
\end{figure*}

In Fig.\ref{fig:withalpha}, we show the ISCO radius (top left), angular momentum (top right), and energy (bottom left) of the particles at ISCO and the energy efficiency, $\eta$ (bottom right) as a function of the MOG field parameter $\alpha$ for the different values of $\beta$ \& $\omega$ parameters. One can see from this figure that an increase of $\alpha$ reduces to increasing $r_{ISCO}$ and ${\cal L}_{ISCO}$ quasi-linearly.

(i) It is also shown that the presence of electric charge and magnetic field interaction parameter $\omega=0.01$ causes to slight decrease in the ISCO radius (see the blue-dashed line). However, the MOG field enhances the effects of the magnetic interaction at higher values of $\alpha$. In the presence of the magnetic dipole of the particle with $\beta=0.1$ the radius slightly increases. 

(ii) The angular momentum of the particles at ISCO increases due to the presence of magnetic interaction force which is centripetal. Similarly, the MOG field enhances the force effect. However, the magnetic interaction with the magnetic dipole of the particles decreases the momentum, which means that in this case the magnetic interaction force is centrifugal. Also, when $\omega<0$ the angular momentum sufficiently decreases. 

(iii) We have found that the energy at ISCO decreases with an increase of $\alpha$, $\beta$ parameters which are positive, and the negative values of $\omega$.

\begin{figure*}[ht!]
\centering
\includegraphics[width=0.42\textwidth]{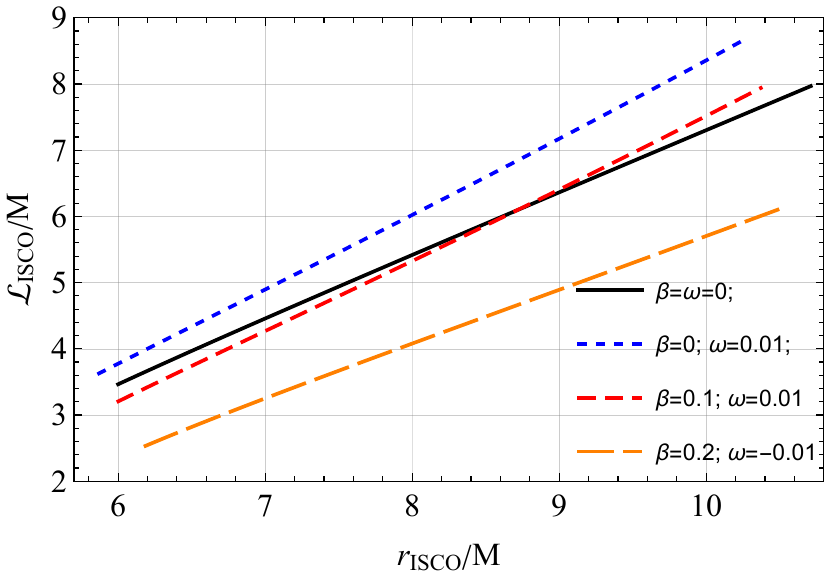}
\includegraphics[width=0.43\textwidth]{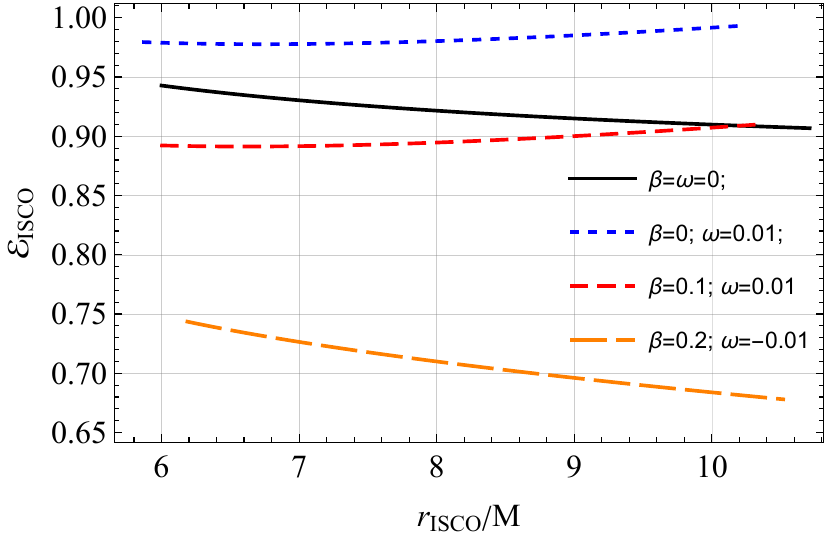}
\includegraphics[width=0.43\textwidth]{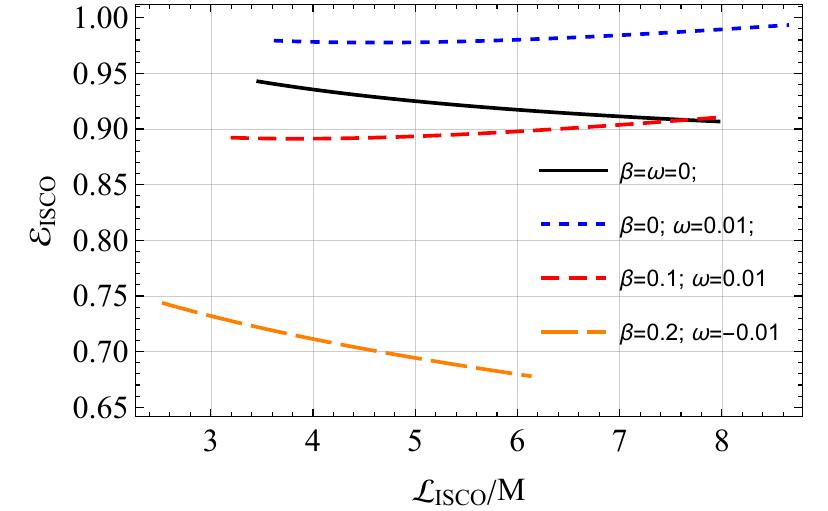}
\includegraphics[width=0.43\textwidth]{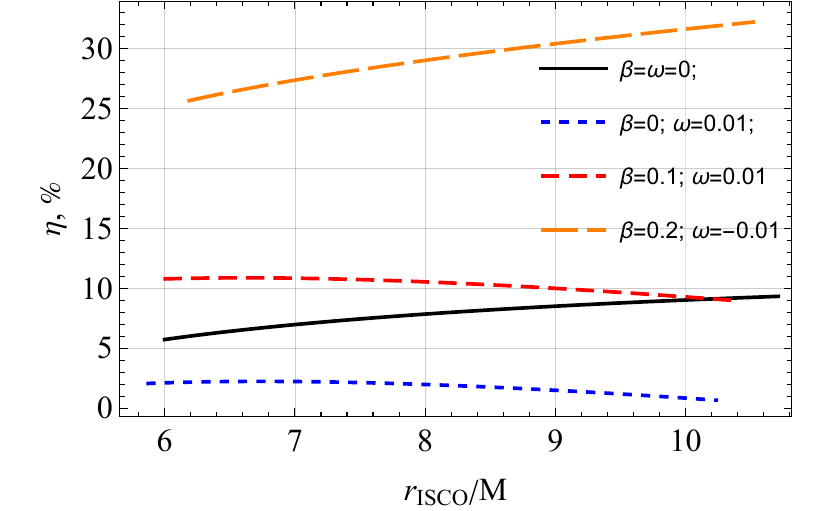}
\caption{The relationships between the ISCO radius and the angular momentum and energy of the particles at ISCO and the energy efficiency.\label{figwithisco}}
\end{figure*}

Figure \ref{figwithisco} we provide detailed analyses of the dependence of the angular momentum and energy of the particles at ISCO and the energy efficiency from the ISCO radius for different values of STVG and magnetic interaction parameters. It is observed from the parametric plots of the ISCO radius and angular momentum and energy of the test particles at the corresponding ISCOs (the energy efficiency) that the $\omega>0$ cases cause to increase (decrease) in the angular momentum (the efficiency). However, in the cases of negative $\omega$ and the presence of $\beta$ cases decrease (increase) them (it).

\begin{figure*}[ht!]
\centering
\includegraphics[width=0.32\textwidth]{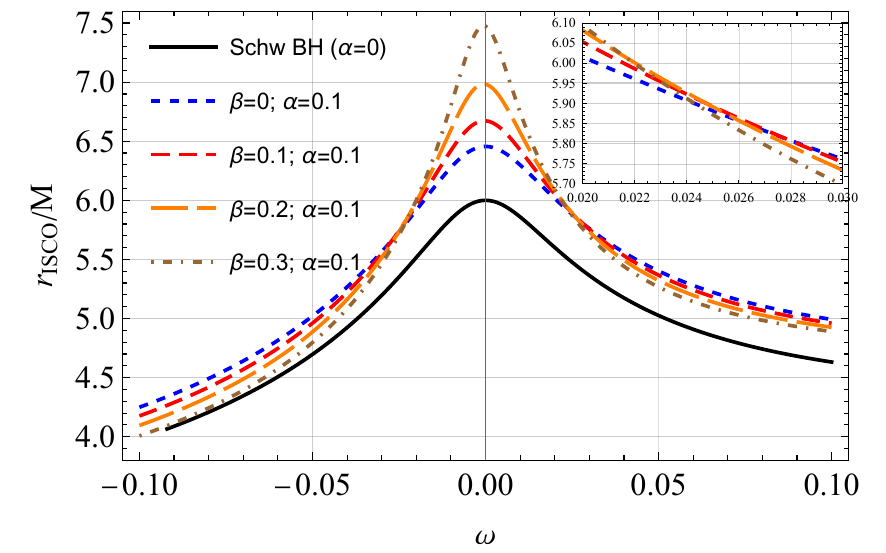}
\includegraphics[width=0.33\textwidth]{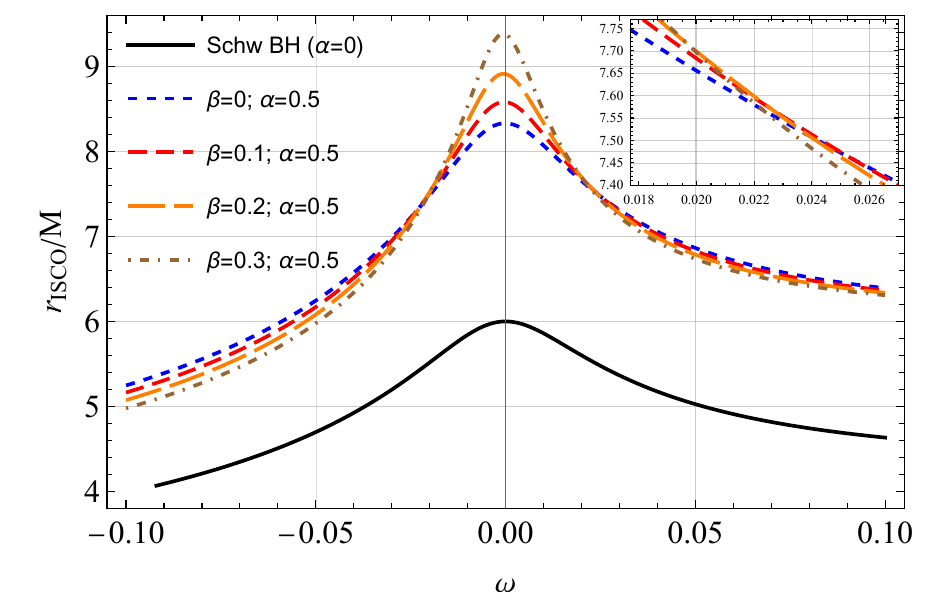}
\includegraphics[width=0.33\textwidth]{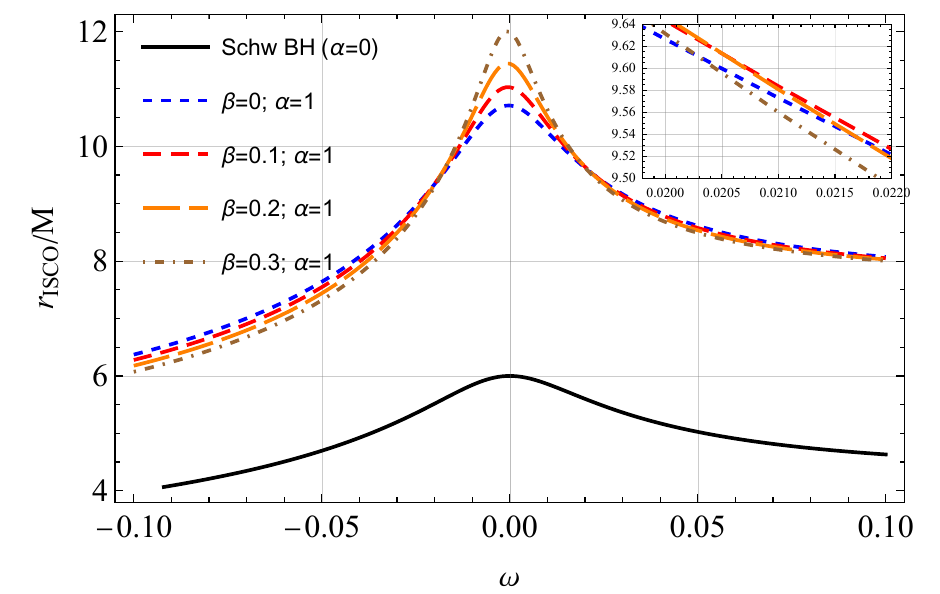}
\includegraphics[width=0.32\textwidth]{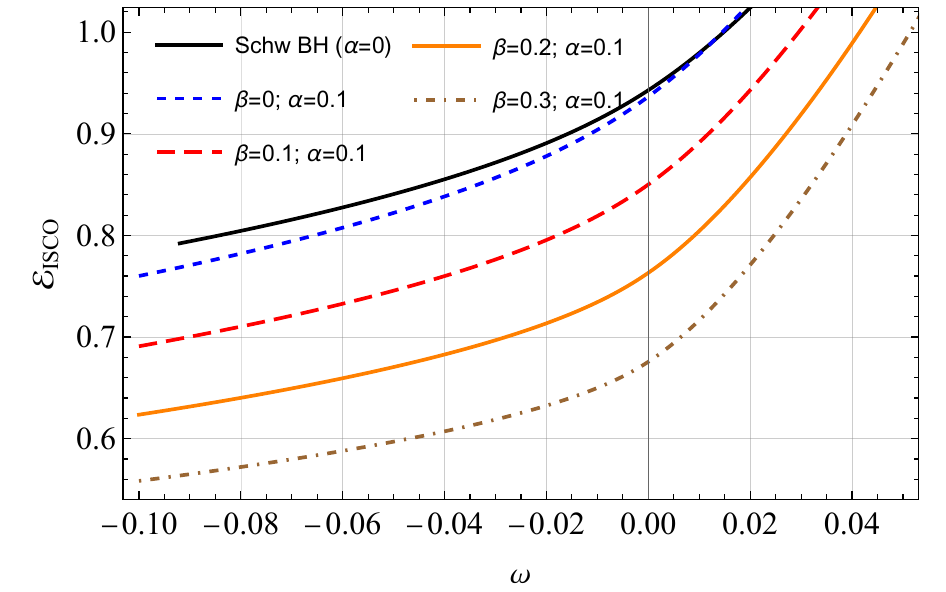}
\includegraphics[width=0.33\textwidth]{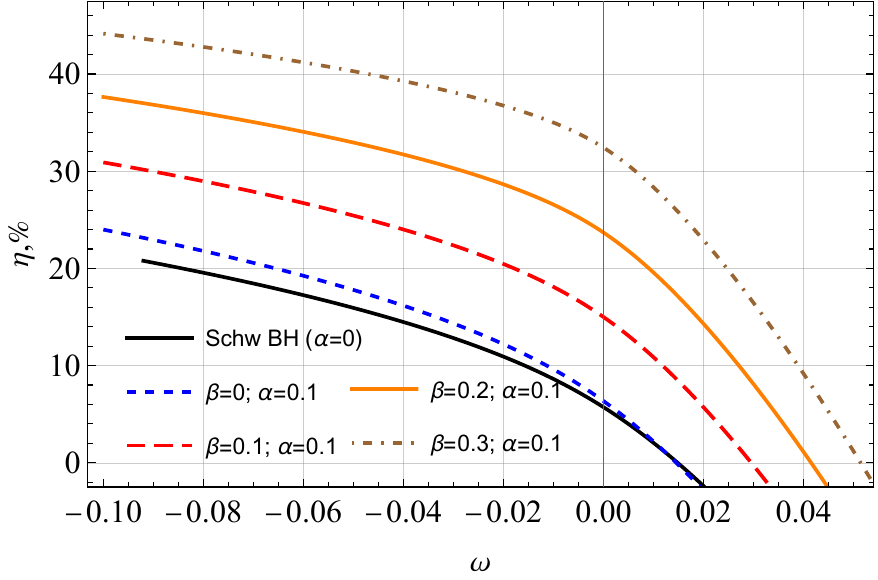}
\includegraphics[width=0.33\textwidth]{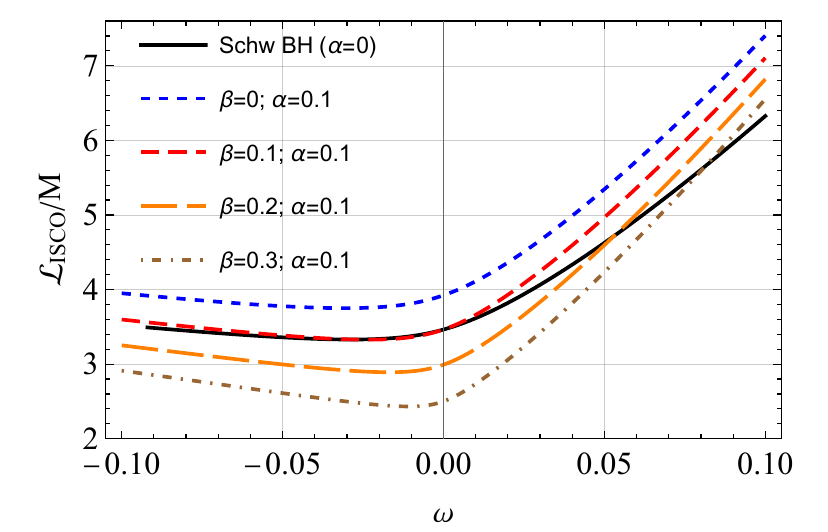}
\caption{The ISCO radius as function of $\omega$ for the different values of $\beta$ and $\alpha$. \label{fig:withomega}}
\end{figure*}

Figure \ref{fig:withomega} shows the dependence of the ISCO radius of the magnetized particles charged in the test on the magnetic coupling parameter $\omega$ for different values of the parameters $\beta$ and $\alpha$. It is also obtained that the increase in the parameters $\alpha$ and $\beta$ causes an increase in the ISCO radius, while both the positive and negative values of $\omega$ reduce it. One can see from the zoomed section of the figure that there is a degeneracy of the different combinations of $\beta$ and $\omega$ that provide the same ISCO at a given $\alpha$. 


\section{Particle collisions near magnetized Schwarzschild MOG black holes  }\label{Sec:Collision}

Estimating the total amount of the energy released by different processes occurring near black holes can explain why the luminosity of AGN is of order $ 10^{45}{\rm erg/sec}$ that may be powered by supermassive black holes.


Several physical models have been proposed as energy extraction mechanisms from black holes. For the first time, Penrose has proposed a simple mechanism ~\cite{Penrose69} by which a particle coming to the ergosphere around a rotating Kerr black hole decays by two particles: one of the parts falls into the black hole, and the other one goes to infinity taking larger energy than the initial one. This mechanism has been developed in works in the literature (for example, \cite{wagh85,Abdujabbarov11,Dadhich18}). 

Banados-Silk-West (BSW) \cite{Banados2009PhRvL,Banados11} have considered collisions of particles near a black hole horizon as an energy extraction model and the model has also been developed in ~\cite{Harada11b,Wei10,Zaslavskii10,Zaslavskii11b,Zaslavskii11c,Kimura11,Banados11,Igata12,Frolov12,Atamurotov13a,Liu11,JuraevaEPJC2021,Tursunov13,Abdujabbarov2020Galax,Stuchlik11a}. It is shown that the energy efficiency extracted from the central black hole is more effective in the cases of head-on collisions.

Here we study collisions of test electrically charged, neutral, and magnetized particles in the spacetime of a magnetized Schwarzschild BH in modified gravity. 
We follow the general expression for the center of mass energy $E_{\rm cm}$ of colliding particles given in Ref.\cite{Banados2009PhRvL}   
\begin{equation}
\left(\frac{1}{\sqrt{-g_{00}}} \ E_{\rm cm}, 0, 0, 0\right)=
m_{1}u_{(1)}^{\mu}+m_{2}u_{(2)}^{\nu},
\end{equation}

where $u_{(i)}^{\mu}$ and $m_{i}$ are the four-velocity and mass of the $i$ th particle. One can obtain the expression for $E_{\rm cm}$ using the normalization condition, $g_{\mu\nu}u^{\mu}u^{\nu}=-1$ in the form

%
%
\begin{equation}\label{ecm2}
\frac{E^2_{cm}}{m_1m_2}=\frac{m_1^2+m_2^2}{m_1m_2}-2g_{\mu\nu} u^{\mu}_1u^{\nu}_2\, .
\end{equation}
In our further analyses, we consider the simple cases in the masses $m_1=m_2=m$.  

\subsection{Critical angular momentum of colliding particles}

In fact, the center of mass energy in the collisions of the particles takes maximum value in close orbits near the horizon,
The radial velocity of colliding particles satisfies the condition $\dot{r}^2\ge0$ that is as a function of the angular momentum $\dot{r}^2(\mathcal{L})$ and the other parameters. As the angular momentum increases, the radial velocity decreases. When the angular momentum takes a critical value the radial velocity and its first derivation along $r$ becomes zero:  $\dot{r}=0$, and  $\partial_r (\dot{r}^2)=0$. We will numerically solve the system of equations in complicated form.

\begin{figure*}[ht!]
\centering
\includegraphics[width=0.49\textwidth]{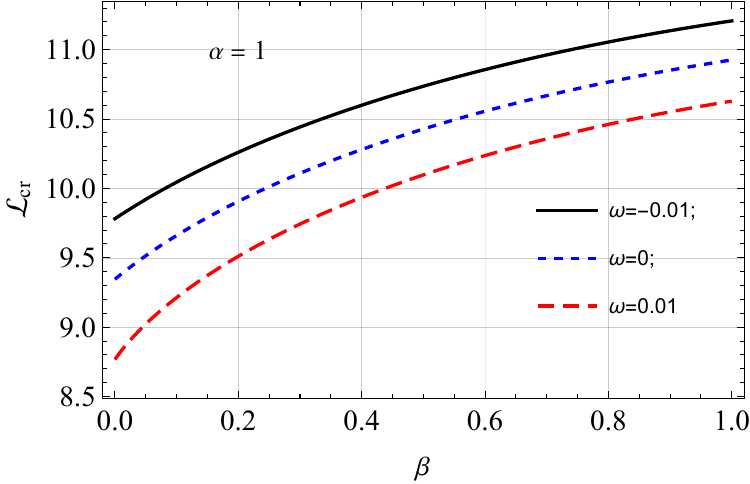}
\includegraphics[width=0.49\textwidth]{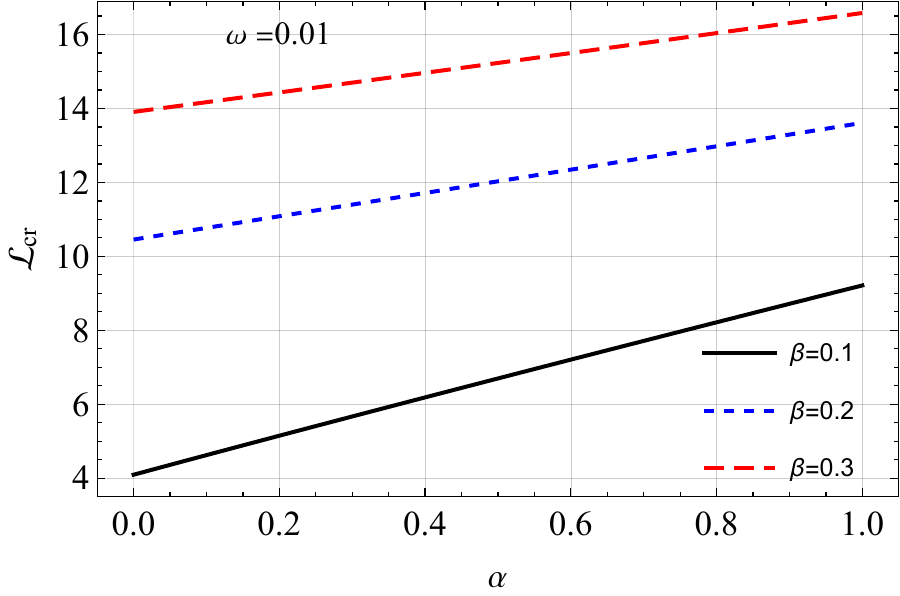}
\caption{The critical value of the angular momentum as a function of $\alpha$ and $\beta$.\label{fig:Lcr}}
\end{figure*}

In Fig.\ref{fig:Lcr} we show the dependence of the critical angular momentum of charged and magnetized particles from the parameters $\alpha$ and $\beta$ for different values of $\omega$. One can see from the figure that negative (positive) values of $\omega$ cause a decrease in the angular momentum. While the MOG field parameter $\beta$ and magnetized one $\beta$ cause increasing it. 

\subsection{Ceneter of mass energy of particles in different scenarios.}

In this subsection, we investigate collisions of different particles: electrically neutral, electrically charged, magnetized particles, and electrically charged particles with magnetic dipole moment. 
The center of mass energy given in Eq.(\ref{ecm2}) takes the following form for the same mass particles:
\begin{equation}
  \frac{E^2_{cm}}{m^2}=2\left(1-g_{\mu\nu} u^{\mu}_1u^{\nu}_2\right)\ .  
\end{equation}
The energy in dimensionless form ${\cal E}^2_{cm}=E^2_{cm}/(2m^2)=1-g_{\mu\nu} u^{\mu}_1u^{\nu}_2$.

\subsubsection{Case I}
First, we consider collisions of neutral particles with (i) neutral, (ii) electrically charged, (iii) magnetized, and (iv) charged-magnetized ones. 

Equations of motion Eqs. (\ref{tdot})-(\ref{rdot}) turns to neutral particles' ones in the case $\beta=\omega=0$.

\begin{figure*}[ht!]
\centering
\includegraphics[width=0.49\textwidth]{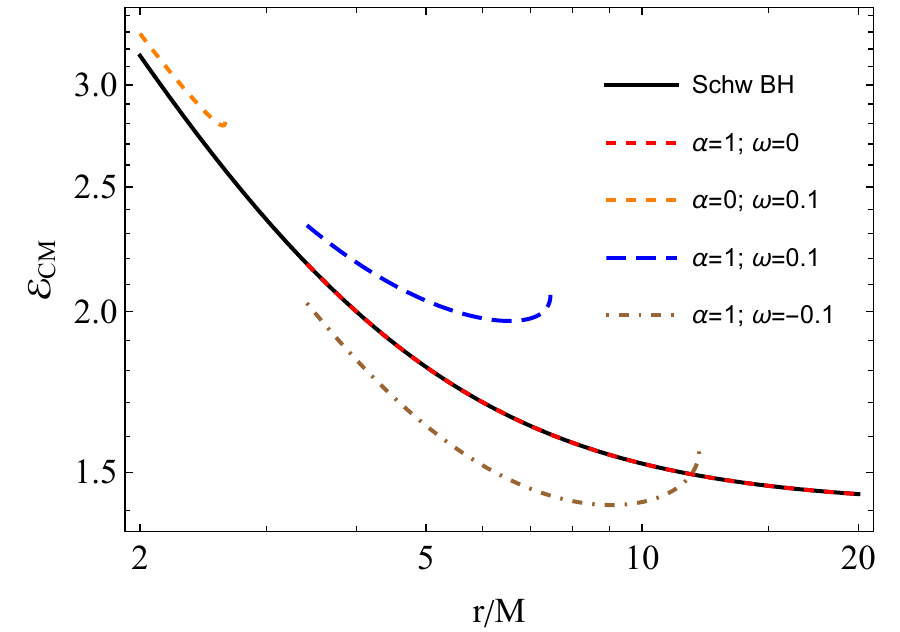}
\includegraphics[width=0.49\textwidth]{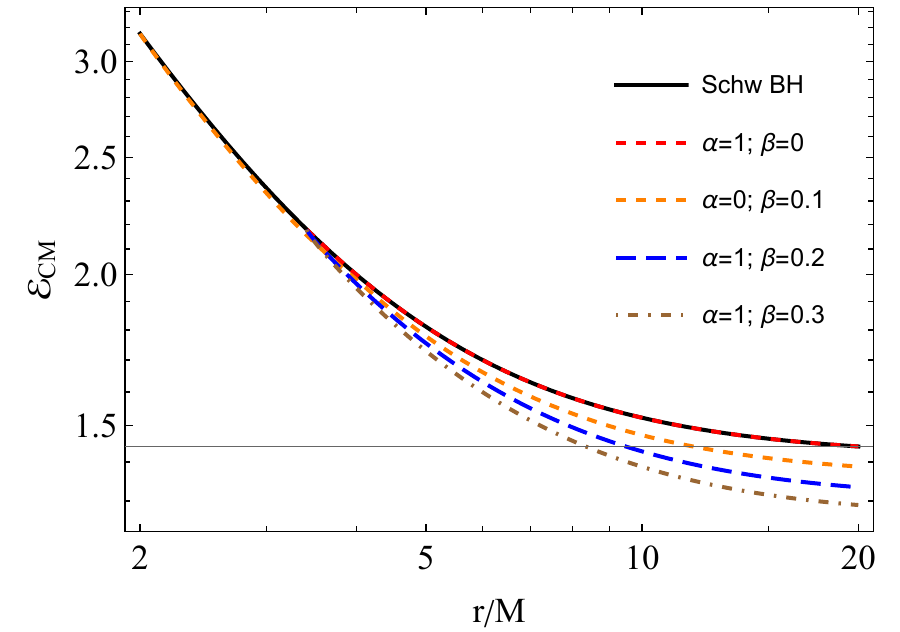}
\includegraphics[width=0.49\textwidth]{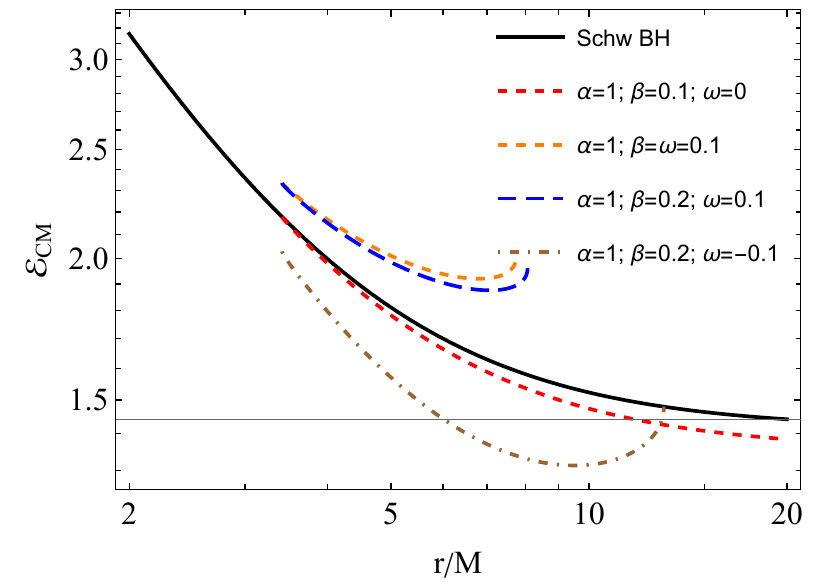}
\caption{Radial dependence of the center of mass energy of collisions of neutral particles with neutral (black and red-dashed lines), electrically charged (top left panel), magnetized (top right panel), and charged-magnetized particles (bottom panel).\label{figNEEcm}}
\end{figure*}

In Fig.\ref{figNEEcm}, we show the radial profiles of the center-of-mass energy in collisions of neutral particles with neutral (black and red-dashed lines), electrically charged (top left panel), magnetized (top right panel), and charged-magnetized particles (bottom panel). The energy is almost the same in MOG at $\alpha=1$ and GR for electrically neutral particle collisions. However, it differs sufficiently for charged particles with $\omega=\pm0.1$. It is observed that in the Schwarzchild case, the collision of charged particles can occur near the horizon (see orange-dashed line in the top left panel) and a bit far from the horizon the collision does not happen due to in repulsive behaviour of Lorentz force. However, the collision occurs at some range far from the horizon of the Schwarzschild MOG black hole with a lower center-of-mass energy than the Schwarzschild black hole in GR when $\omega>0$. In contrast, for negative $\omega$ the energy is sufficiently small from the GR case. Meanwhile, in the case of neutral particle collisions ${\cal E}_{cm}$ is the same as in GR. Also, an increase in $\beta$ causes a decrease in energy.

\subsubsection{Case II}

Here, we consider collisions of electrically magnetized particles with (i) neutral, (ii) electrically charged, and (iii) charged-magnetized particles.

\begin{figure}[ht!]
\centering
\includegraphics[width=0.49\textwidth]{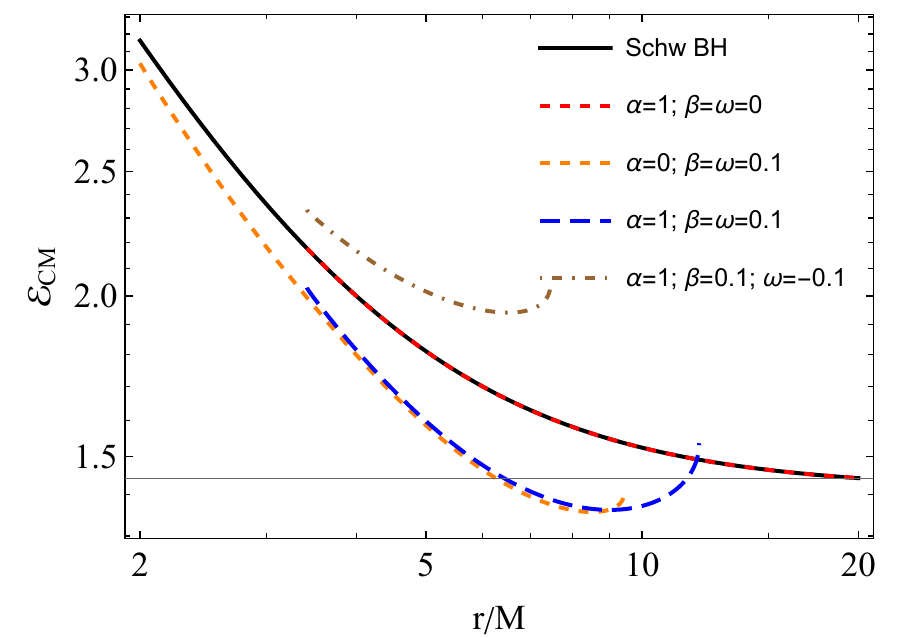}
\includegraphics[width=0.49\textwidth]{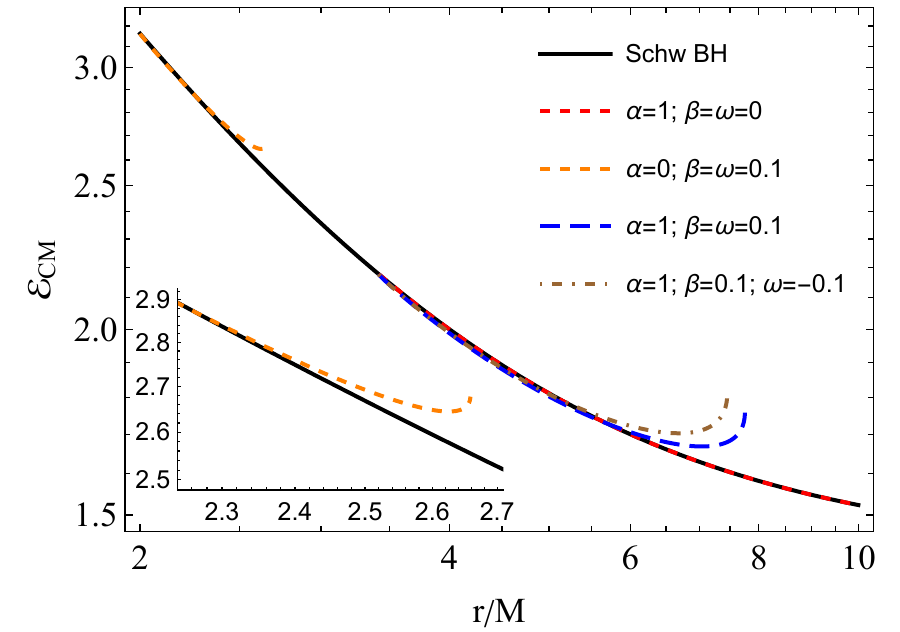}
\caption{The same figure with Fig.\ref{figNEEcm}, but for the collisions of magnetized particles with electrically charged particles (left panel) and electrically charged particles with charged-magnetized particles (right panel).\label{figEMEcm}}
\end{figure}

In Fig.\ref{figEMEcm} we show the radial behavior of the center-of-mass energy in the collision of magnetized particles with electrically charged particles (left panel) and the collisions of electrically charged particles and charged magnetized particles (right panel). In this figure, we consider the magnetic interaction parameters of both particles to be the same, that is, the particles have a magnetic dipole moment, and the second one has an electric charge that satisfies the same value of the magnetic interactions $\omega=\beta$. Similar effects of $\beta$ on the center of mass energy can be observed as we have seen in Fig.\ref{figNEEcm}. However, in the collisions of magnetized and charged particles case, the combined effects of $\omega$ and $\beta$ of the particles differ from the previous one. In other words, in the case when the second particle has a negative charge $\omega$, the energy increases with respect to the energy of the collisions of neutral particles. While the energy decreased by the effect of negatively charged particle collisions. Moreover, the energy has a minimum at a point from where it starts to increase again in which the magnetic interaction energies increase radially, and then the energy curve disappears due to the dominant effect of magnetic interactions. The distance where the energy takes minimum increases with the growth of $\alpha$ and $\omega$. In the case of electrically charged and charged-magnetized particle collisions, the magnetic interaction effects on the center of mass energy are weaker due to the combined effects of MOG and magnetic interactions with electric charge and magnetic dipole moment. As usual, collisions between charged and magnetized particles do not occur at far distances under the effects of repulsive Lorentz forces. The distance is also dependent on $\alpha$, for higher values of $\alpha$ the distance goes far, similarly, it also shifts slightly out for the particles which have higher $\omega$ and $\beta$.

\subsubsection{Case III}

Here, we consider collisions of magnetized particles with (i) neutral, (ii) charged-magnetized, and (iii) magnetized particles.

\begin{figure}[ht!]
\centering
\includegraphics[width=0.492\textwidth]{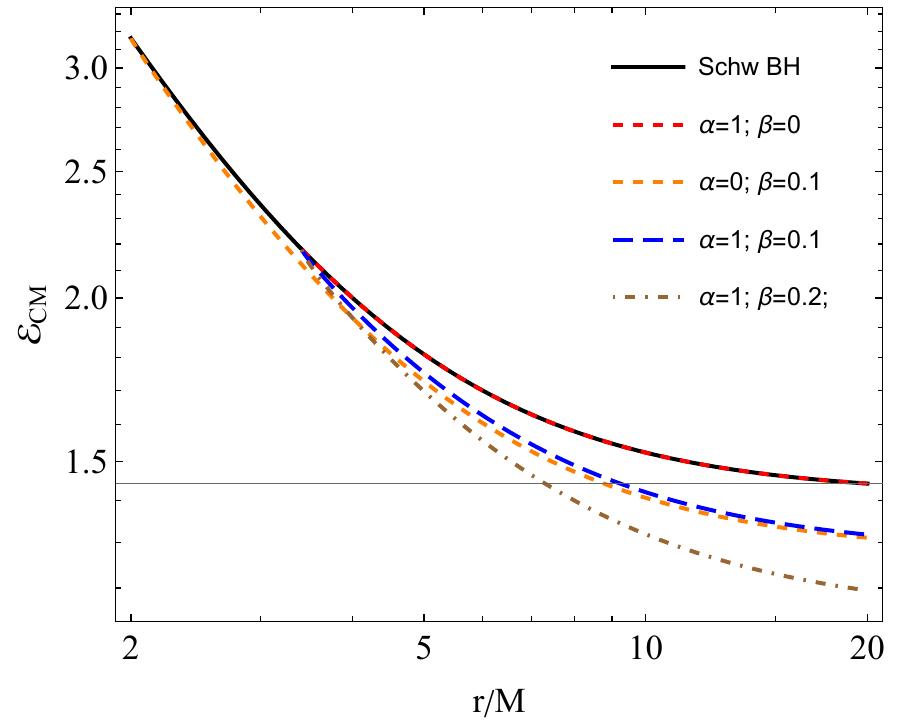}
\includegraphics[width=0.492\textwidth]{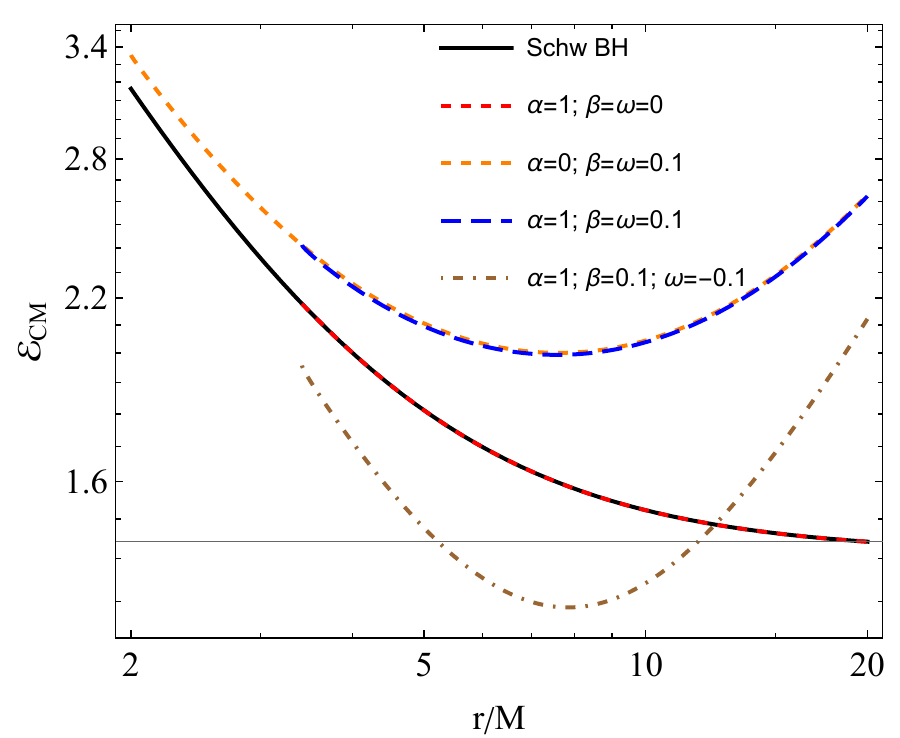}
\caption{The same figure with Fig.\ref{figNEEcm}, but for the collision of two magnetized particles (left panel) and magnetized particles with charged-magnetized particles (right panel).\label{figMEMEcm}}
\end{figure}

It is observed that the center of mass energy decreases slightly as both the MOG field and magnetic interaction parameters increase in the magnetized-magnetized particle collisions (see left panel of Fig.\ref{figMEMEcm}). In this figure, we take the same value for the magnetic interaction parameters of the two colliding particles as $\beta=0.1$. Also, it is seen that ${\cal E}_{cm}$ increases (decreases) when the value of $\omega$ is positive (negative).

\section{Conclusion} \label{sec:6}
In the present work, we have investigated the effects of (electro)magnetic interactions between test charged particles and magnetized particles on the dynamics of particles in strong gravitational fields around Schwarzschild black holes in modified gravity established by J. Moffat \cite{Moffat06}. We have studied the dynamics of particles having electric charge and magnetic dipole moment in the spacetime of Schwarzschild-MOG black holes.
Also, we have provided a solution of Maxwell equations for the angular component of electromagnetic four potentials in the Schwarzschild-MOG spacetime.

We have obtained equations of motion and effective potential for circular motion of such particles using a hybrid form of the Hamilton-Jacobi equation which includes both interactions of electric charge and magnetic dipole moment with the external magnetic field assumed as asymptotically uniform and interaction between the particles and the MOG field. We have obtained that the effective potential for neutral particles in the presence of an MOG field decreases, while at large distances the potential matches with the GR case, which means that the MOG field effects in Moffat's model disappear at large distances. Also, the MOG field effects enhance the effects of magnetic interactions on the effective potential. 

Also, we have studied behaviors of ISCOs radius and the energy \& angular momentum of charged and magnetized particles at ISCOs together with the energy efficiency under the effects of magnetic and MOG field interactions. It is obtained that an increase of $\alpha$ reduces to increasing $r_{ISCO}$ and ${\cal L}_{ISCO}$ quasi-linearly and the presence of electric charge and magnetic field interaction parameter $\omega$ causes a slight decrease in the ISCO radius. Also, it is shown that the MOG field interaction enhances the effects of the magnetic interaction on ISCO. While, the presence of the magnetic dipole of the particle the radius slightly increases. The angular momentum of the particles at ISCO increases in the presence of centripetal magnetic interaction forces. Similarly, the MOG field enhances the interaction force effect. However, the magnetic interaction with the magnetic dipole of the particles decreases the momentum, which means that in this case, the magnetic interaction force is centrifugal. Also, when $\omega<0$ the angular momentum sufficiently decreases. The energy in ISCO decreases with an increase in the parameters $\alpha$, $\beta$, which are positive values and negative values of $\omega$.

Finally, we provide detailed analyses of the effects of the three interactions mentioned above on the center of mass energy in the collisions between neutral, electrically charged, and magnetized particles.
    
It is shown that the collision of charged particles can occur near the horizon and far from the horizon the collision does not happen due to in repulsive behaviour of Lorentz force. However, the collision occurs in some range far from the horizon of the Schwarzschild MOG black hole with a lower center of mass energy than the Schwarzschild black hole in GR when $\omega>0$. In contrast, for negative $\omega$ the energy is sufficiently small from the GR case. Meanwhile, in the case of neutral particle collisions ${\cal E}_{cm}$ is the same as GR.

 We have fixed the magnetic interaction parameters of both particles to be the same $\omega=\beta$.  In the collisions of magnetized and charged particles case, it is obtained that the combined effects of $\omega$ and $\beta$ of the particles differ from the previous one. 
 
 Moreover, the energy has a minimum at a point from where it starts to increase again in which the magnetic interaction energies increase radially, and then the energy curve disappears due to the dominant effect of magnetic interactions. The distance where the energy takes the minimum increases with the growth of $\alpha$ and $\omega$. In the case of electrically charged and charged-magnetized particle collisions, the magnetic interaction effects on the center of mass energy are weaker because of the combined effects of MOG and magnetic interactions with electric charge and magnetic dipole moment. As usual, collisions between charged and magnetized particles do not occur at far distances under the effects of repulsive Lorentz forces. The distance is also dependent on $\alpha$, for higher values of $\alpha$ the distance goes far, similarly, it also shifts slightly out for the particles which have higher $\omega$ and $\beta$. 
\section{acknowledgement}

This research is supported by Grant No. FA-F-2021-510 of the Uzbekistan Agency for Innovative Development. F.A., J.R., and A.A. acknowledge the ERASMUS+ ICM project for supporting their stay at the Silesian University in Opava. S.M. gratefully acknowledges support from Grant FZ-20200929385 of the Ministry of Higher Education, Science, and Innovation of the Republic of Uzbekistan.

\bibliographystyle{apsrev4-1}
\bibliography{references,reference}

\end{document}